# Cavity receiver designs for parabolic trough collector

Khaled Mohamad[a], P. Ferrer[a,b]

[a]School of Physics, Materials for Energy Research Group, University of the Witwatersrand, Johannesburg, 2001, South Africa.

[b]School of Physics, Mandelstam institute for Theoretical Physics, University of the Witwatersrand, Johannesburg, 2001, South Africa.

Solar parabolic trough collector technology has received significant attention among researchers for its economic potential to meet electrical and thermal energy demands. Furthermore, parabolic trough collector power plants can be hybridized with fossil fuels or other renewable power plants. Improving the efficiency of the parabolic trough collector leads to increase the electricity production and allows the parabolic trough collector power plant to be built in a more compact size, at reduced capital investment. This review focuses on parabolic trough collector receivers, which uses cavity technology to improve efficiency. The cavity receiver has shown potential to overcome shortages of the conventional receiver. It was seen that the thermal losses of the cavity receiver are affected by various parameters such as aperture sizes, emissivities, and working fluid temperature as well as mass flow rate and wind velocity. Different cavity designs have a different effect on the solar flux distribution along the circumference of the receiver. An efficient cavity receiver still has the potential to minimize the losses and hence, improve the overall efficiency. Some of the design gaps have been identified as a guide for future work.

## 1. Introduction

Among the concentrated solar power technologies, the Parabolic Trough Collector (PTC) is the earliest and most widely accepted design for harnessing solar energy for the dispatchable source (it refers to the source of energy that can be used on demand or at the request of power grid operators) of electricity or industrial and chemical, due to its low cost and reliability [1]. In this technology, solar radiation is focused onto a focal line aimed at the receiver unit. The receiver heats up and in turn, imparts a significant portion of its heat to a Heat Transfer Fluid (HTF) circulating within. The hot HTF can be used to generate electricity through a steam cycle or in thermochemical applications. The receiver unit is one of the most complex parts of the Parabolic Trough Collector (PTC). Its optical and thermal properties directly influence the performance of the entire system, and the efficiency of the whole system largely depends on it [2]. Therefore, the receiver unit has to be carefully designed in such a way to minimize energy losses.

The conventional type of the PTC receiver unit for PTC is evacuated glass metal. It consists of a blackened absorber pipe encapsulated by the glass cover. An evacuated region in between minimizes convective heat losses, see Figure 1.

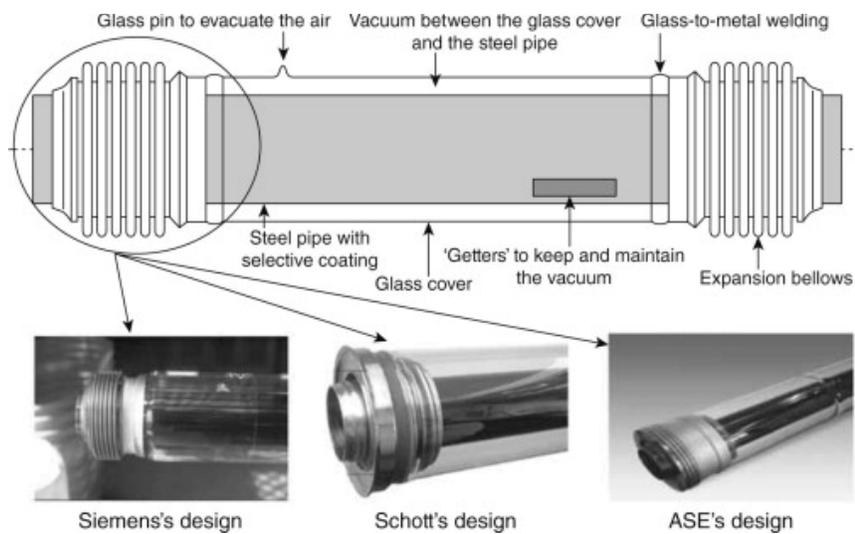

Figure 1: Structure of a parabolic trough receiver [3].

The thermal contacts between the receiver pipe and the glass cover are kept to a minimum to reduce conduction heat losses. The conventional receiver has advantages such as high efficiency and low heat loss. However, it has some deficiencies, such as:

- The breakage of the glass and the metal glass seal, which leads to vacuum loss in the annular region between the absorber pipe and the glass cover [4].
- The permeation of hydrogen gas into the annular region is due to the thermal decomposition of the organic heat transfer fluids (HTF) such as Therminol VP-1. The hydrogen gas in the annular region is found to increase the heat losses to four-time those in the evacuated receiver [5].

- Lower efficiency at high temperatures [6][7].
- Non-uniform thermal distribution and thermal stress problems [7][8].
- High capital outlay and high maintenance cost [9][10].

Increasing the outlet HTF temperature is one of the main challenges to improve the overall efficiency of the PTC power plant and to reduce the Levelized Cost of energy and the combined thermal storage. For example, thermal storage quantity can be reduced to one third, if the HTF temperature increases from 350 °C to 600 °C [5]. Currently, most of the power plants used oil-based fluids, which operate to a limited temperature up to 400 °C. The dominant heat losses mechanism at high temperatures are due to the thermal emission (IR) from the receiver pipe. This loss is conventionally minimized by painting the receiver pipe with a spectrally selective coating. It is a material that absorbs well in the visible region of the solar spectrum and emits poorly in the IR region. Much work has been published on selective coatings and improving their properties [11]. Dudley et al. [12] investigated the performance of black chrome and cermet selective coatings, and found that the Cermet had lower emissivity values and thus reduced thermal losses and improved efficiency. Forristal [13] compared six different selective coating materials to evaluate receiver unit performance. He showed that the receiver performance is susceptible to the optical properties of the selective coatings, and improving the coatings could result in significant efficiency gain. Cheryl et al.[14] Investigated spectrally selective coating materials for concentrated solar power applications, and concluded that the ideal selective coating material should be easy to manufacture, low-cost, chemically and thermally stable in air at operating temperature of 500 °C. Because the selective coatings are placed on absorber pipe of the receiver, the receiver operation temperature is limited and hence thermal efficiency. Kennedy et al. [15] were able to successfully model a solar-selective coating composed of materials stable to 500 °C using computer-aided design software. Archimede Solar Energy (ASE) [16] manufactured the world's most advanced solar receiver tube with selective coating. It operates at temperature up to 580 °C with molten salts as HTF.

Further, the solar receiver SCHOTT PTR 70 is designed for solar thermal power plants operating with oil-based HTF at a temperature up to 400 °C [17]. The heat loss measurements for SCHOTT PTR 70 are carried out in a round-robin test performed by SCHOTT Solar in cooperation with Deutsches Zentrum für Luft-und Raumfahrt (DLR) and US National Renewable Energy Laboratory (NREL). The tests confirmed that the heat loss is less than 250 W/m at working temperature 400 °C [17].

Some complications start to appear at a high temperature (> 300 °C) and with a non-uniform solar flux distribution such as the thermal performance and thermal stress of the receiver unit. When a non-uniform heat flux profile is incident on the receiver unit, the temperature across the circumference varies, and peaks/hot spots in the receiver start to increase with temperature. It leads to bending of the absorber pipe and breaking of the glass cover. P Wang et al. [8] found that the maximum temperature gradient for the safe operation of receiver tubes is about 50 K. These complications have been addressed to increase the life span of the receive absorber. Some recent research focused on improving both thermal transfer and uniformity of the thermal distribution [7], but sacrifice pressure drop of the receiver unit or increase the quality of the absorber pipe and other components, adding cost [6].

Some of these studies suggest applying inserts into the absorber pipe such as metal foam, porous discs, perforated plates or coiled wire turbulators inserts. A metal foam inserted into the absorber pipe facing the concentrates sunlight reduces the thermal stress, decreases the temperature difference on the outer surface of the absorber pipe by about 45%, but increases flow resistance [8]. Experimental [18] and theoretical work [19] was conducted for the porous disc insert application (a disc perforated with holes inserted into the pipe), increasing the thermal efficiency between 1.2% and 8% according to the numerical study [20]. The coiled wire turbulators insert application has been examined experimentally and numerically [21]. At the pitch distance 30 mm of coiled wire turbulator (a coil-shaped wire inserted into the absorber pipe), the heat transfer enhancement is approximately twice that of the smooth tube [21].

Furthermore, Eliamasa-ard, Thianpong and Eiamsa-ard [22] studied the effects of three different twisted tapes inserts on thermal enhancement, Nusselt number (Nu) and friction factor ($f$), which are a single twisted tape, twin-counter (co-twisted) tape and counter (co-swirled) tape. The experiments of four different twist ratios (2.5, 3, 3.5, and 4) were conducted. The twist ratio is the ratio of the axial length for 180° turn and the inner diameter of the absorber tube. The results showed that the thermal enhancement, Nu and $f$ were increased with decreased the twist ratio, and co-swirled tape has significant heat transfer enhancement compared to co-twisted tape. Other research works investigated the heat transfer performance and pressure drops of the absorber pipe with wire-coil inserts. It was found that wire coil inserts increased the turbulence inside the absorber pipe and Nu increased to 330% [23][24]. In addition, the effect of the wire-coil inserts material types (aluminum, stainless steel and copper) on $f$ and the heat transfer rate were studied by Shashank and Choudhari [25]. They found that the copper wire coil inserts have improved the heat transfer rate compared to stainless steel and aluminum.

Nanan, Thianpong and Pimsarn [26] conducted experimental and numerical studies to investigate tube inserts with baffle turbulators (straight baffles, straight cross-baffles, straight alternate-baffles, twisted baffles, alternate twisted-baffles and twisted cross-baffles). The twisted cross-baffles showed the highest thermal performance, while the straight cross-baffles showed the lowest thermal performance. Yuxiang Hong, Du J and Wang S [27] performed experimental investigation of the thermal and fluid flow characteristics of a spiral grooved tube (SGT) fitted with twin overlapped twisted tapes in counter large/small combinations (TOTT-CL/S). The experiments were conducted for plain tube (PT), SGT and SGT fitted with TOTT-CL/S. The heat transfer rate of SGT fitted with TOTT-CL/S was higher than that of the SGT and PT but larger pressure drop.

Other studies focus on geometrical structure improvement for the absorber pipe of the receiver unit such as a dimpled tube, unilateral milt-longitudinal vortex-enhanced tube as well as symmetric and asymmetric outward convex corrugated tubes. These inserts manipulate the Reynolds number, substantially improving the "mixing" of different temperature layers of fluid. A numerical study showed improved performance of a dimpled absorber pipe under non-uniform heat flux over uniform heat flux [28]. Similar improvements were found for the unilateral milt-longitudinal vortex-enhanced tube, with better heat transfer performance than the smooth pipe under a wide range of working conditions [29]. The introduction of symmetric [30] and asymmetric [31] outward convex corrugated tubes, regular outwards "bulges" in the absorber pipe, effectively decrease the thermal strain and enhance the heat transfer performance.

Further, researchers have investigated Nanofluids (with suspended nanoparticles) to enhance the heat transfer. The most used nanofluids contain nanoparticles such as Al, $Fe_2O_3$, $Al_2O_3$, Cu, $TiO_2$, and $SiO_2$ [32][33]. Nanofluids tend to have more significant thermal conductivity than normal heat transfer fluids. The thermal conductivity increases by decreasing the particle size and increasing the volume fraction and temperature [33]. E. Bellos found that the use of the nanofluids increases the efficiency of the collector by 4.25% [32].

Some of the limitations associated with evacuated receivers have been overcome. Nevertheless, the monopoly over this technology and the cost of the receivers put barriers for solar projects, especially in the developing countries. Accordingly, finding alternatives to evacuated receivers that can compete in terms of efficiency and cost-effectiveness is a prime target. Such alternatives are expected to have a significant impact on the already long-standing industry.

An alternative to the evacuated receiver is the cavity receiver, where the concept of the cavity receiver comes from the blackbody principle. The blackbody is an ideal object, absorbs all incident radiation, regardless of direction and wavelength [34]. The object that most closely resembles a blackbody is a large cavity with a small opening. The radiation that is incident through the opening has very little chance to escape, it is either absorbed or undergoes multiple reflections before being absorbed [34]. The cavity receiver that can achieve the blackbody principle could make up the shortages of the evacuated receiver, as mentioned earlier. Although early studies of the cavity receiver lacked detail in terms of optimizing, efficiency, and thermal losses, the recent researches are looking to improve and come up with innovative designs. In this paper, we review different cavity receiver designs. This review is intended for the receiver of a parabolic trough collector, but some designs are suitable for the receiver of the linear Fresnel reflector systems. This work aims to study different cavity designs. It is worthwhile to gather innovative ideas that help to enhance the overall efficiency and identify the gaps in these designs for future research work. In the following discussion, we divided the cavity receiver designs into cylindrical receiver units and non-cylindrical receiver units.

## 2. Cylindrical receivers

The base design of the cavity receiver consists of a cylindrical metal tube with a cavity opening for the incident solar radiation. In most of the cavity receivers that have been studied, the space inside the cavity was at atmospheric pressure. The inner cavity surface opposite to the aperture window was mirrored. The aperture window of the cavity was at the bottom, facing the parabolic mirror collector and it was closed by a transparent cover to reduce convection and re-radiation heat loss. Moreover, the outer cavity surface was thermally insulated to minimize the effect of heat loss by convection and radiation to surroundings.

**A. Mirrored glass cover with uncoated aperture**

Ramchandra et al. [5] studied and optimized the cavity receiver shown in Figure 2 for minimum heat loss. The study intended to evaluate and compare different PTC cavity receiver alternatives using a validated numerical model. Their cavity receiver, shown in Figure 2, consists of a mirrored cavity surface facing the absorber pipe. The cavity aperture is closed by a transparent glass cover to reduce re-radiation and convection heat losses. The annulus between them is separated by an air gap at atmospheric pressure. An annular ring of microtherm between the absorber and glass envelope at both ends is placed to suspend the absorber [5].

The dimension of the cavity aperture was carefully selected alongside with the focal line of the collector to ensure that all reflected radiation enters the cavity through the aperture and to minimize inaccuracies in both tracking and directional errors due to the suns shape. The optimum annulus dimension between the absorber pipe and the cavity envelope was selected to reduce convection and conduction heat loss. The optimum dimension of the annulus largely depended on the diameter of the receiver components [35].

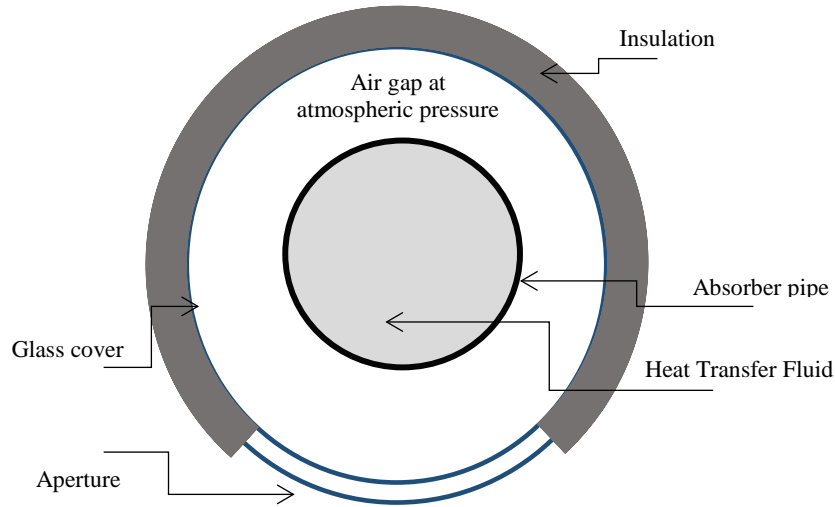

Figure 2: Schematic representation of the linear cavity receiver[5].

Although this cavity receiver with a parabolic trough of small rim angle (around 45°) was capable of being a suitable alternative to the conventional receiver, it had some limitations. First, the intercept factor (the incident energy that enters the aperture and reaches the absorber) was lower than that of a conventional receiver. This problem was addressed by coupling this cavity design with a lower rim angle (around 45°) for the parabolic trough. The conventional receiver had a maximum concentration at rim angle = 90°, whereas the concentration ratio for this design was maximum at rim angle = 45° [5]. Second, the selective coating failed because the air in the annular gap oxidized the coating. The oxidation of the selective coating in the presence of the air in the annulus was expected to be resolved with the progress in developing selective coatings [36]. Third, at a higher temperature, the thermal conductivity of the insulation material increased, thus the heat loss also increased [35]. This issue was solved by selecting a better-suited insulation material.

**B. Hot-mirror coated glass cover**

A similar philosophy to the above mention design by Ramchandra et al. [5] of a cavity receiver that evolved from applying a reflective mirror coating on the inner side of the glass cover to trap the radiation inside the receiver. Ferrer et al. [37] [38] suggested using a hot mirror coating, a dielectric material that is transparent to the visible region of the solar spectrum and reflected well in the IR region. The study intended to reduce the heat losses, especially at a high temperature and reduce the thermal stress of the glass cover as well as introducing an alternative to the conventional receiver with a selective coating. In this suggested design, the annulus between the absorber pipe and the coated glass cover was under vacuum, see Figure 3. Hot mirror coating was first implemented for energy-efficient windows in automobiles and buildings [39] and for applications related to concentrating photovoltaics and thermophotovoltaics [40][41]. There are two general types of hot mirror films: a semiconducting oxide with a high doping level and a very thin metal film sandwiched between two dielectric layers (see [42][43][39] for more details). The thin metal film coating shows some unavoidable losses, while the highly doped semiconducting oxide shows more advantages, i.e., Indium-Tin-Oxide (ITO). The hot mirror coating for a solar collector must meet some performance specifications. It needs to be highly transparent in the visible region and have high reflectivity in the IR region of the solar spectrum. Granqvist et al. [46] and Lampert et al. [44] focused on improving the transparency in the visible region and the reflectivity in the IR.

For PTCs, the effects of the hot mirror film have been modeled and studied previously, see Figure 3. Grena [43] simulated the system, including heat reflection using hot mirror films with simplifying assumptions, and his results showed an increase in overall efficiency tested over a year by 4%. Also, a 2D simulation in this regard showed the possibility of increasing the working fluid's temperature to over 400 °C [45]. Other efforts of this type used a three-dimensional model to take into account the radiation exchange by using different segmented surfaces inside the receiver along the pipe's length. This study showed that the hot mirror receiver effectively reduced the IR losses at higher temperatures, reduced the thermal stress on the glass cover, and suggested the use in a hybrid system [37].

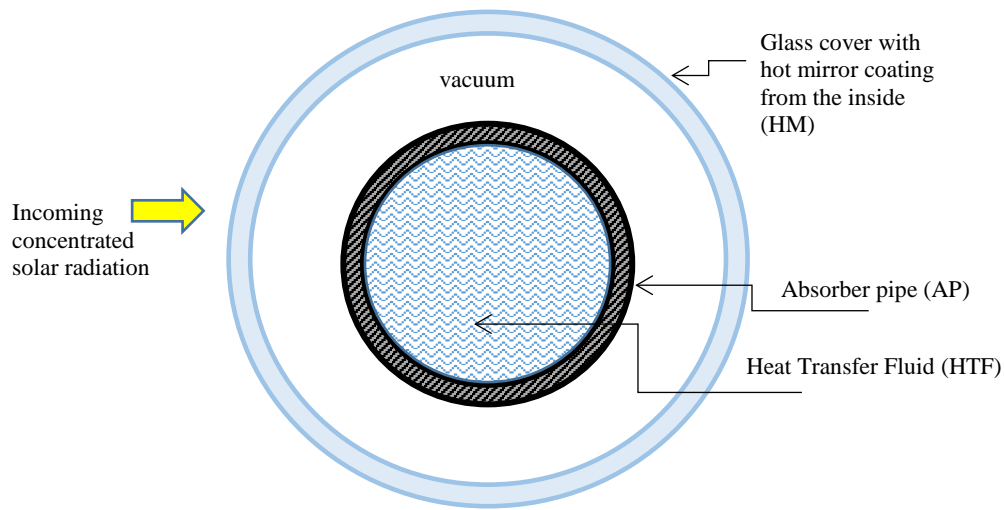

Figure 3: The cross-section of the receiver unit with a hot mirror coating on the glass cove envelop [37].

Practically, current hot mirror coatings have not been found to improve on the overall efficiency of solar plants, but have the capacity of reaching much higher temperatures than their selective coating counterpart. Hot mirror type systems are currently investigated, which may, in the near future, surpass selective coating technology [37]. If a hot mirror coating was found with better visible transmission/IR reflection properties than a selective coating, then the hot mirror coating would have some advantages.

**C. Mirrored cavity receiver with a hot mirror application on the aperture**

K Mohamad and P Ferrer filed a patent on a cavity receiver for PTC [46][6]. This cavity receiver design combined the use of the mirrored cavity receiver with a hot mirror coating in a novel way [6]. The design aims to make up the shortages of the conventional receiver and achieve higher efficiency, higher HTF temperature, and significantly cut the costs per unit PTC. The receiver consists of a highly reflective hot mirror coating on the inside of the borosilicate glass cover on the cavity aperture. The inner cavity surface was coated with a highly IR reflective material, such as polished aluminum. A vacuum in between minimized convective losses see Figure 4. The highly polished inner cavity surface reflects thermal radiation onto the absorber much more effectively than the hot mirror coating. The authors showed that the design was able to achieve the highest thermal and optical efficiencies, especially if the hot mirror coating has a higher transmission for the incident solar radiation. Novel aspects of the background theory for this design related to IR reflectivity in the receiver annulus were added and implemented in a simulation code [6]. The model of the IR reflectivity inside the receiver annulus was validated in [47][38]. The simulation results indicated that the proposed design could exceed the HTF temperature ceiling compared to existing alternatives and could hence potentially increase the efficiency of the system. Further, the cavity geometry and a hot mirror coating at the aperture enable increased retention of thermal radiation of the receiver.

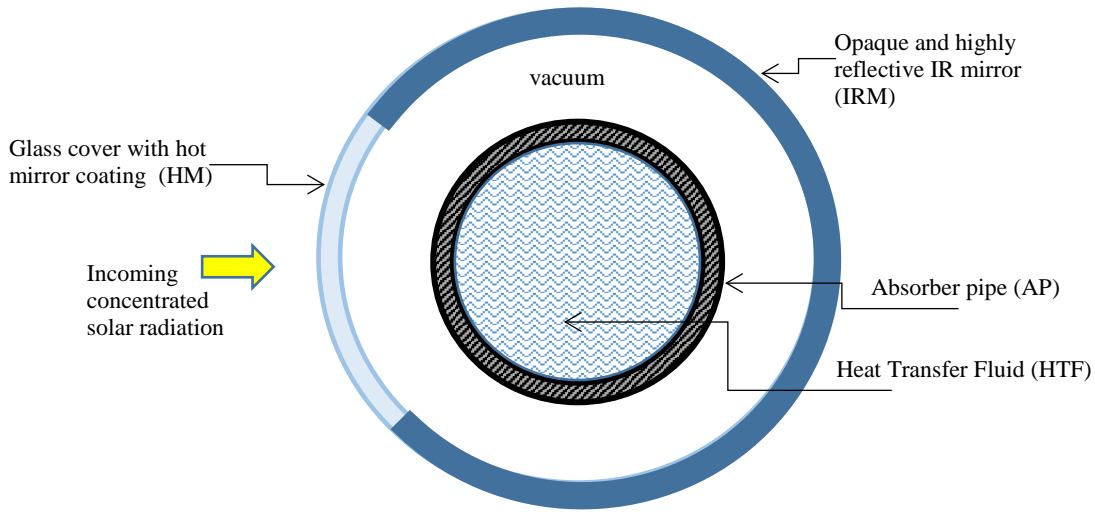

Figure 4: the cross-section of the cavity receiver with a hot mirror coating on the cavity aperture [48].

The cavity system was simulated and studied in many aspects using the operating conditions and design parameters for SEGS (Solar Electric Generating System) LS2. It is one of three generations of parabolic troughs installed in the nine SEGS power plants in California [12]. Different aperture sizes were simulated. The smallest aperture reached the highest temperature and thermal efficiency. The maximum HTF temperature for this design rose close to 1300 K if the aperture opening size was 50% of the total size of the outer receiver circumference and 1490 K if the aperture opening size was 15% of the outer receiver circumference size. The receiver efficiency of the smallest aperture was at 33% compared to 27% for the largest opening, at receiver length of 400 m. Furthermore, different reflectivities for the inner surface of the cavity were studied (92%, 95%, and 98%). It was seen that the highest reflectivity reached the highest temperature, at receiver length of 500 m (about 200 K higher than the lowest reflectivity), and had the best receiver efficiency (33% compared to 27% for the lowest, at receiver length of 400 m). The effect of the hot mirror coating was studied by comparing the system with and without the hot mirror application. At the aperture opening size of 15% of the total size of the outer receiver circumference size and the reflectivity of the inner cavity mirror 98%, the stagnation temperature was found to be ~250 K higher with coating. The cavity system without hot mirror coating initially performed better at a lower temperature, but after 775 K, the hot mirror coated system dominated performance.

Furthermore, a comparison between this cavity design (in case of a cavity opening with a hot mirror coating, an opening size of 15% of the total outer receiver circumference, and the cavity mirror reflectivity of 98%) and conventional receivers with a selective coating and without coating (bare) was conducted. The results indicated that the selective coating performed slightly better in terms of efficiency at a lower temperature, but the cavity system dominated at higher temperatures, see Figure 5 and Figure 6.

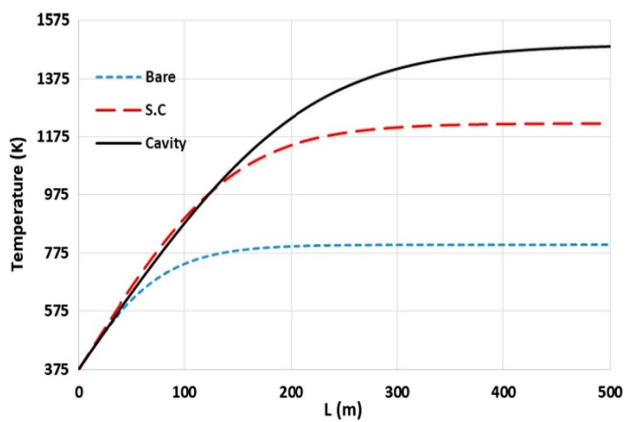

Figure 5: HTF outlet temp for 375 K inlet temperature of different designs [6].

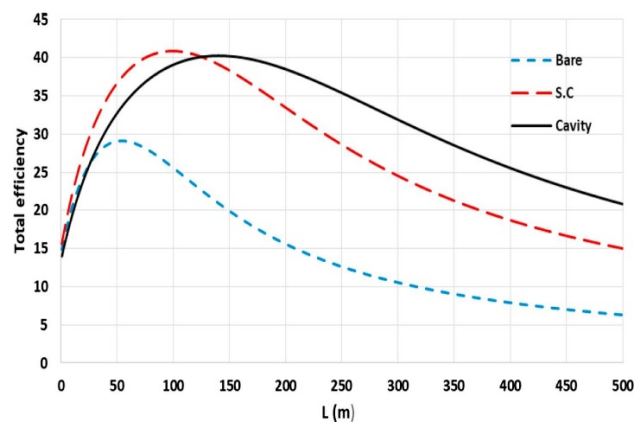

Figure 6: The total efficiency of the system as a function of the length of the receiver with different designs [6].

Although the cavity system seemed theoretically capable of reaching very high temperatures, it is unlikely that it will be used as such. A very high reflectivity (> 99%) inside the cavity walls, which is well within the realm of possibility, will make the system very efficient. Higher efficiency will likely allow the HTF to achieve optimum plant temperature in a shorter length, thus reducing the needed AP length. This system is likely to reduce costs and the overall size of the system [6].

**D. Cavity receiver with asymmetric compound parabolic concentrator**

Roman Bader et al. [49] has proposed a cavity receiver, presented in Figure 7, with the aim to significantly cut the costs per unit PTC through a decrease in the PTC size and use of low-cost materials as well as using the air as the heat transfer fluid. The cavity design consists of an absorber tube enclosed by an insulated stainless steel cylindrical cavity with an asymmetric compound parabolic concentrator (CPC) at the cavity aperture, show in Figure 7. This design was tested with a 43 m long prototype installed and a 9 m aperture solar trough concentrator to study the efficiency of receiver under different operating conditions and validate the heat transfer model of the receiver that based on Monte Carlo ray-tracing and finite-volume methods [49].

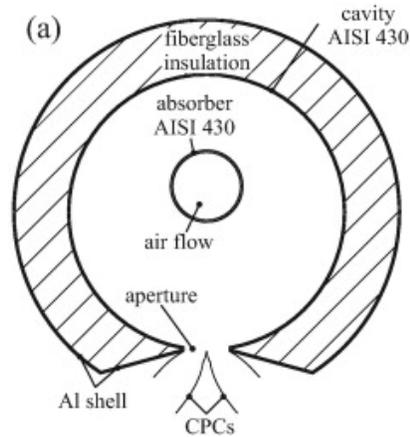

Figure 7: Schematic representation of the cavity receiver design proposed by[49]

At HTF inlet temperature of 120 °C, HTF outlet temperature was expected in the range of 250-450 °C, at summer solstice solar noon (input of 280 kW), while the efficiency of the receiver ranged from 45% to 29%. The optical losses were one-third of the total incident radiation on the receiver, due to the spillage at the aperture and reflection inside the cavity. The thermal losses due to the natural convection from the cavity insulator were 5.6 - 9.1%. Another heat loss was due to the re-radiation through the cavity aperture (6.1 - 17.6%). Moreover, HTF pumping work had an associated energy penalty of 0.6 – 24.4% of the total power generated [50].

The authors suggested a modified cavity receiver design that evolved from the initial design, which is discussed next.

**E. Corrugated cavity receiver**

Roman Bader et al. [49] suggested modifications to their previous cavity design. The modified design is shown in Figure 8. It consists of a cylindrical cavity with a smooth or corrugated black inner surface with a single or double-glazed aperture window. The modification can be summarized as follows:

The absorber pipe was eliminated to allow the HTF (air) to flow in a sufficiently large cross-section through the cavity to compensate for the lower volumetric heat capacity. The cavity from the inside (surface 1, Figure 8) was enhanced with V-corrugations to increase the heat transfer surface area. The cavity aperture was made from glass to reduce the convective heat loss at the aperture, where the glass is almost opaque for the radiation emitted from a blackbody at < 600 °C [49].

The two window panes of a double-glazed window with air in between were used to trap the emitted radiation of surface 1 and the inner window to reduce the heat conduction through the window.

The authors studied this design with four different receiver configurations: smooth, V-corrugated absorber tube, single, and double-glazed aperture window.

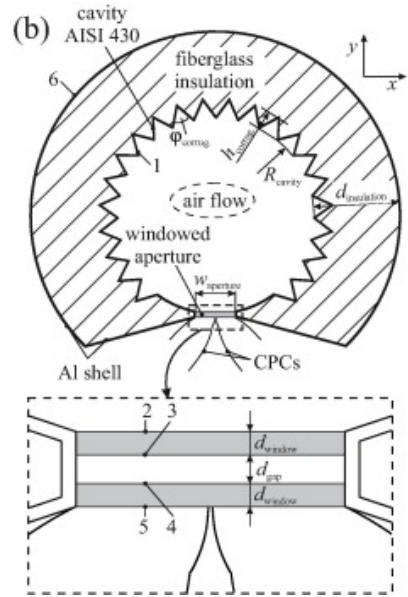

Figure 8: The modified design [42].

Using the air as a HTF in this design is to avoid the chemical instability during the operation and to allow the direct coupling of the solar collector with a packed bed thermal storage [51][52]. However, the air has a lower volumetric heat capacity, which leads to lower convective heat transfer compared to conventional HTFs, such as molten salt and thermal oils. This problem could be solved by designing a receiver with a larger diameter and higher heat transfer area than that of the conventional receiver [53]. In the modified design, V- corrugations were used to increase the heat transfer of the surface area.

The simulation was validated with the experimental work for the design in Figure 7, and then used to analyze the modified design. The modified design Figure 8 was simulated during the summer solstice at solar noon in Sevilla, Spain, with direct normal solar irradiance equals 847 W/m$^2$. It had a collector efficiencies between 60% and 65% at HTF temperature of 125 °C and between 37% and 43% at 500 °C. The largest source of energy loss was the optical loss, which was more than 30% of the incident solar radiation. It was mainly because of the absorption by the concentrators and reflection at the receiver's aperture window [49]. Moreover, the required pumping power for the HTF through a 200 m long receiver operated between 300 °C and 500 °C was between 11 and 17 kW.

Furthermore, increasing the HTF mass flow rate would cause a decrease in heat loss with the corrugated absorber pipe, which led to an increase in receiver efficiency. At the expense of additional reflection loss, the double-glazed aperture window significantly reduced the re-radiation loss from the receiver's aperture compared to the single-glazed window, where the double-glazed acted as an effective radiation trap. However, the single-glazed receiver led to higher collector efficiencies at low HTF temperature (< 300 °C), the double-glazed receiver led to higher collector efficiency at high HTF temperatures. The author suggested using a material with a high reflectivity on the aperture window to improve the overall efficiency of the PTC [49].

**F. Cavity receiver with copper pipes and copper annulus**

Barra et al. [54] proposed a different design of the cavity receiver shown in Figure 9. The motivation of their study was to design a blackbody cavity receiver more efficient and less expensive than the conventional receiver units. The structure of that cavity receiver was made from iron and copper pipes, which were selected for economic and stability reasons. The iron oxidation was to induce high visible absorption in the cavity. The V-shape Pyrex glass in the cavity opening was removable and selected for low-cost commercial availability and to reduce the convection loss. This design was tested with 50 m$^2$ parabolic trough prototype, and the experiment parameters and results used to build a mathematical model to simulate the system. The working conditions of this cavity receiver were not optimum, but the performance of that design appeared promising compared to those obtained from a more expensive receiver. The study showed less solar interception with the cavity and more thermal losses, which were strongly depending on wind intensity and direction. Furthermore, the cavity receiver without a vacuum and selective treatments reached a good performance, but still inferior to the current receivers with highly selective coating and vacuum. The author suggested that the cavity receiver needed a design improvement and proposed the substitution of copper pipes with a copper annulus, as shown in Figure 10. This design was first proposed by Boyd et al. [55] to achieve better performance without requiring advanced materials and coating. The inside of the annular tube was coated with black paint. The entrance aperture with the insulation on the wall side was cut in V- shape to limit the thermal radiation loss from the collector aperture. It was assumed to have a diffuse surface to reduce the effect of radiative field view of the aperture to the surrounding [55]. In addition, the natural convection loss could be restricted by adjusting the entrance angle of the aperture because the narrow passage causes flow restrictions

[55]. The receiver efficiency has been calculated for the hot end of the receiver, where it is lowest, and also averaged over the length of the collector, as a measure of overall performance [55]. The hot end efficiency was 53%, and the average was estimated to be 76% at 370 °C. Furthermore, the conduction through the insulator was dominant at a lower temperature (at 130 °C) and radiation loss was dominant at a higher temperature (at 370 °C).

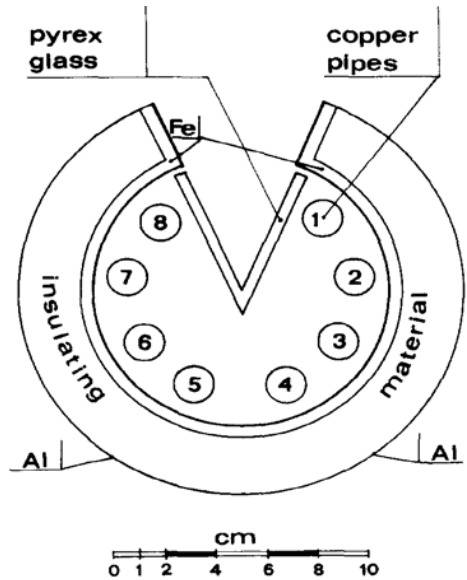

Figure 9: Vertical cross-section of the cavity receiver[54].

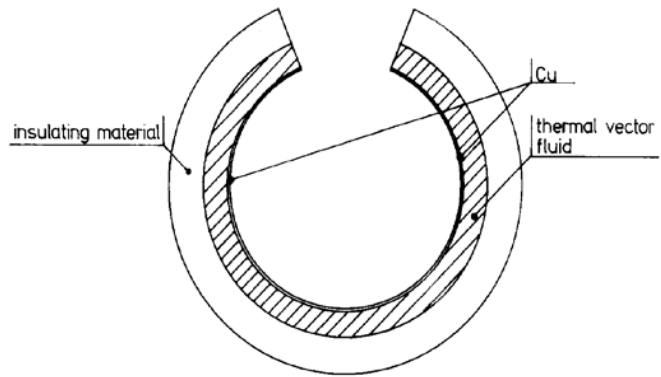

Figure 10: Vertical cross-section of the suggested optimized cavity receiver[54].

## G. The arc-shaped cavity receiver

Xueling Li et al. [56] studied an arc-shaped linear cavity receiver, where the absorber has a crescent shaped channel, as seen in Figure 11, with the aims to raise the HTF temperature and reducing the cost of production and maintenance as well as having a similar shape and size to the conventional receiver [56]. The crescent shaped channel was made from a copper and the outer surface of the channel covered with opaque insulation. The cavity aperture window was fabricated from borosilicate glass, see Figure 11.

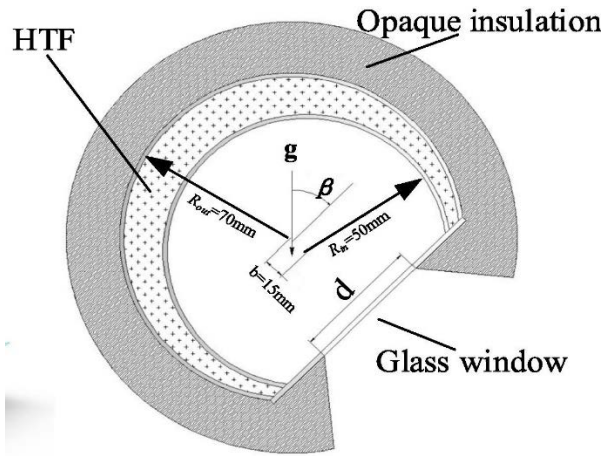

Figure 11: Arc-shaped linear cavity receiver with a lunate channel [56].

The thermal performance of this design was studied theoretically using a numerical model [56]. The effects of the HTF temperature, surface emissivity, inclination angle, and aperture width were analyzed and displayed some of the following characteristics: the total heat loss of the receiver decreased from 394.5 W/m to 335.8 W/m, when the inclination angles increased from 0° to 90°. The heat loss of the receiver increased with the aperture width of the cavity at the same collecting temperature. A reasonable aperture width of this cavity design is about 50 - 70 mm, where the dimensions of the cavity are shown in Figure 11. Generally, at a larger aperture width, the heat loss increased, and the optical loss (which could be from installation and tracking errors, mirror roughness and manufacturing error) decreased. At a high temperature ( > 400 K), a comparison between the proposed design and an evacuated receiver (Solel's UVAC) showed that the heat loss of the proposed design was less and slower than that of the evacuated receiver [56].

## 3. Non-cylindrical receiver unit geometries

### A. Elliptical cavity receiver

Fei Cao et al. [57] studied the elliptical cavity receiver, as shown in Figure 12, with the aim of having a receiver with high thermal performance, low cost, and not frangible. The outer cover was elliptic with an open inlet towards the parabolic trough mirror. The incoming concentrated solar radiation was incident on the receiver and entered into the absorber pipe through the opening.

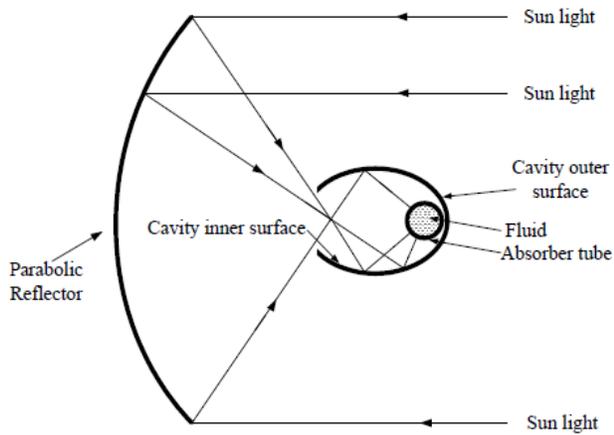 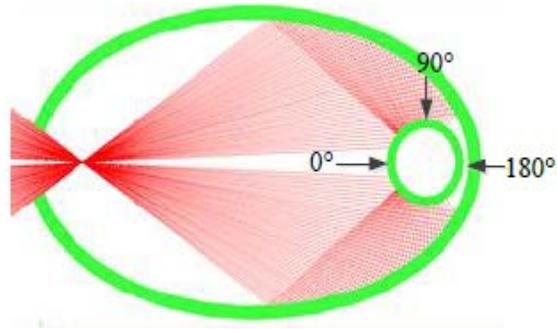

Figure 12: Elliptical cavity receiver [57].　　　　　Figure 13: Heat and light flux distributions around the receiver cavity [57].

The position of the absorber pipe was on the opposite focus from the opening. The design was studied using a ray-tracing model and a heat transfer model. The study focused on analyzing the heat transfer and the heat flux distribution in a 1m cavity receiver tube as well as the thermal stress distributions of the absorber pipe. The geometry of the design and its dimension were mentioned in [57]. The light and heat flux distribution around the absorber pipe outer surface is summarized in Figure 13. There was no light and heat flux from 0° to 10° and from 125° to 215°, see Figure 13. Most of the other light entered the cavity and reflected from the elliptic inner surface to the absorber surface. This lead to a non-uniform heat flux absorption along the circumference of the absorber tube. The thermal stress was found to cause a maximum deformation of the receiver tube along the fluid direction of 3.1 mm at 0.82 m. The thermal stress was generated by the temperature difference between the tube inner/outer wall and the heat transfer fluid pressure and phase. This, in turn was caused by the fluid and steel characteristic, solar heating flux and the characteristics of the heat transfer fluid and the absorber material. Besides that, equivalent stress along the absorber pipe was found at higher fluid mass flow rates.

**B. Elliptical cavity receiver with optical funnel**

Fei Cao et al. [58] proposed a modified design that evolved from their initial design in Figure 12 to improve the performance of the elliptical cavity. The authors modified the design by adding a flat plate reflector at the cavity aperture, see Figure 14, to diminish the effect of changing the PTC focal length on the cavity opening length and the effect of tracking error angle on the cavity performance.

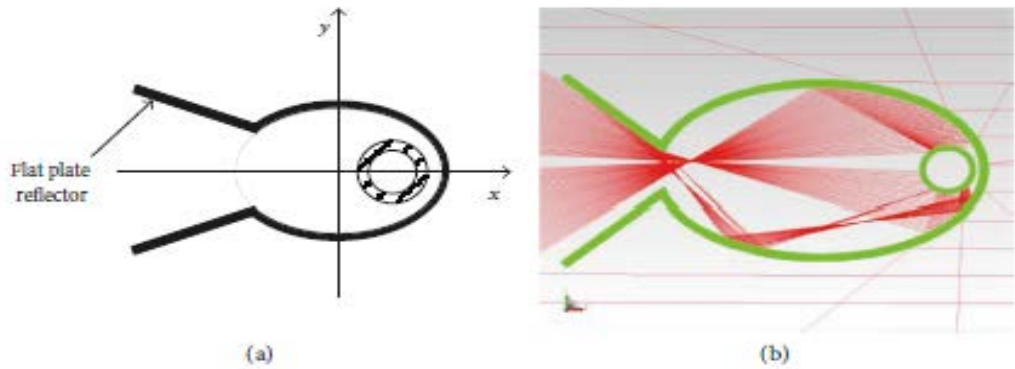

Figure 14: Elliptical cavity with flat plate reflector [58]   a) Structure schematic    b) light distribution around the receiver.

Figure 15 shows the effects of different tracking error angles and PTC focal distances on the cavity darkness of the modified cavity receiver. Cavity darkness is the percentage of the sunlight on the absorber surface to the total incident sunlight. It was found that introducing the flat plate reflector can significantly increases the cavity darkness, where the flat plate breaks the monotonic relationship of the cavity darkness under different focal distances. Further, at different tracking error angles the incident sun radiation is reflected by the flat plate, which causes different multi-reflections inside the cavity, which leads to the curves in Figure 15 [58].

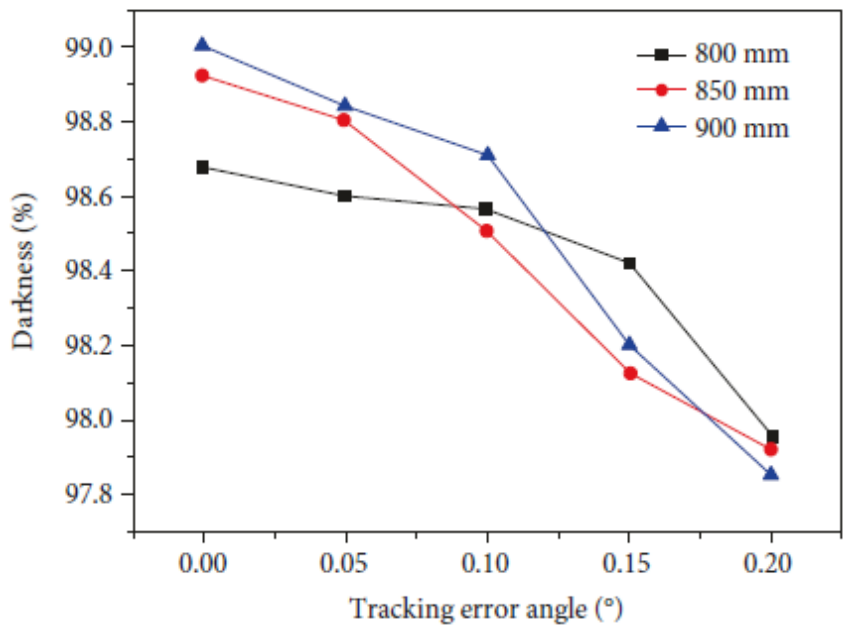

Figure 15: Darkness of the modified elliptical cavity receiver under different tracking angle and PTC focal distances [58].

## C. Triangular and V cavity receivers

F Chen et al. [59] studied the triangle cavity receiver experimentally and theoretically with the aims to provide high efficiency and make up the shortages of the conventional receivers such as high cost, leakage during long-term running, and challenging technology. The cavity was made from aluminum, and the absorber surface was a triangle or V-shape structure with fins in the dorsal side to enhance the thermal performance between the absorber surface and the HTF, see Figure 16.

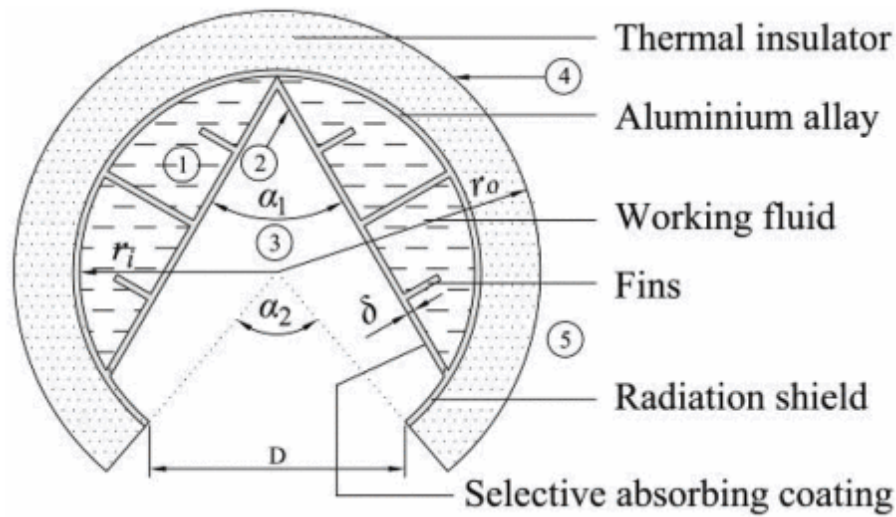

Figure 16: Cross-section of the triangle cavity receiver [59].

A radiation shield was placed at the aperture to reduce the view factor of the absorber pipe to the surrounding ambient. Experimental and theoretical work was conducted to study the heat loss due to the absorber inclination and the ambient wind speed. The results showed that the ambient wind speed had a relatively significant effect on heat loss, while the impact of the absorber inclination was relatively small. The heat loss was found to be 51.2 W/m, 53.2 W/m, 59.7 W/m and 61.8 W/m, when the inclination angles were 0°, 30°, 60° and 90°, respectively, at temperature differences between the working fluid and ambient of 150 ± 3 °C. At the inclination of 60° and temperature differences of 27 °C, 84.5 °C, 126.7 °C and 176.3 °C, the heat losses were 11.5 W/m, 30.4 W/m, 48.25 W/m and 71.55 W/m, respectively.

Furthermore, the heat losses of the cavity receiver were 76.6 W/m, 85.5 W/m, 96 W/m and 102.6 W/m, when the wind speeds were 1 m/s, 2 m/s, 3 m/s and 4 m/s, respectively, at the temperature difference of 150 ± 3 °C. At the wind speed of 3 m/s the temperature differences were 27.6 °C, 85.6 °C, 124.4 °C and 172.6 °C, and the heat losses were 17.3 W/m, 63 W/m, 83.85 W/m and 104 W/m, respectively.

In a windless case, the heat loss of this design was equivalent to that of UVAC3 evacuated receiver and the new-generation (UVAC2008)[60].

Two studies investigated the same design with an additional glass cover on the aperture [61][62]. Their objective was to reduce the heat loss from the aperture. The first study focused on heat transfer performance, where the investigation was theoretically and experimentally. The optical performance was studied using Monte Carlo ray-tracing method [61]. The design showed high optical efficiency of about 99% because the concentrated sunlight repeatedly reflected by the triangular shape with almost no escape [61]. Furthermore, the heat flux distribution of the heating surface of the design was heterogeneous, which could cause thermal stress at higher temperature [61]. The second study focused on thermal performance, also from a theoretical and experimental perspective [62]. The design had a good thermal performance in the medium temperature range, and it was comparable to that of the evacuated tube in that temperature range. The study involved the effect of the glass cover and the fins. These additions showed an improvement in thermal performance. Moreover, the heat transfer fluid temperature in this cavity design could exceed 570 K [62].

Fei Chen et al. [2] studied the optical properties of the triangular cavity absorber using a theoretical method. They found that the cavity absorber's aperture width, depth to width of the triangular shape, and the offset distance from the focus of the triangular cavity were important parameters to improve the optical performance, where the cavity optical efficiency was 89.23%. It was recommended to select the depth to width ratio of 0.8 to 1, the aperture width of 70 mm, and the offset distance of 15 mm [2].

Figure 17 shows another design that is similar to the V- shape or triangular cavity. It consists of a center tube as absorber and two inclined fins which acted as the inner cavity surface with the glass cover over the aperture. A rectangular shell separated the cavity from the surrounding. The space between the shell and the absorber was filled with aluminum silicate fiber and asbestos rubber sheet between the glass cover and the end of the inclined fins. At the sides of the shell, there was a fixed axle to make the whole cavity movable. It rotated the system when solar irradiance was high enough to collect the concentrated solar energy [63][64]. The movable cavity mechanism was a novel design to prevent overheating while reducing heat loss. The study was based on experimental and theoretical investigations.

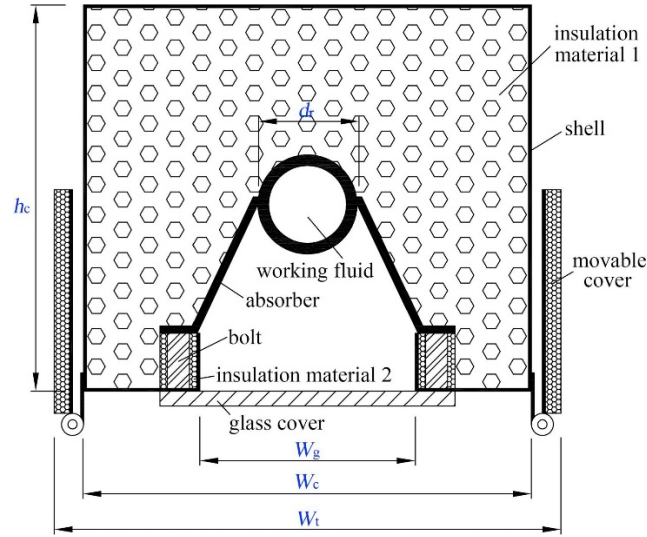

Figure 17: Cross-section of the V-shape cavity receiver [63].

Regarding the effect of the on-off state of the movable mechanism of the cavity on the heat loss, it was found that the heat loss of turning off the movable cover was less than that of turning on [64], where the heat loss reduction rate varied from 6.36% to 13.55%. The author found that the movable mechanism should be optimized for thermal insulation performance and operation control strategy [64]. The collector efficiency was tested at different inlet temperatures ranging from 80.6 °C to 160.5 °C and the mass flow rate from 170 to 181 g/s. The efficiency was in the range of 34.2% to 48.5% [63]. Further, the thermal conductivity of insulation materials was significantly improved the thermal performance of the design, i.e., the collector efficiency increased by 1.47 times, if the thermal conductivity of the insulation materials changed from 0.1 to 0.02 W. K/m [63]. Besides, the collector efficiency could increase by decreasing the emittance of the absorber and glass cover, i.e., the collector efficiency increased from 34.45% to 38.49% if the emittance of the absorber $\varepsilon_{ab}$ and the glass cover $\varepsilon_G$ drop down from 0.9 and 0.95 to 0.1 and 0.15, respectively.

The efficiency of this design was comparable to the efficiency of the metal glass evacuated tube (64.25%). If we use the following optimized parameters:

- $\varepsilon_{ab} = 0.15$,
- $\varepsilon_G = 0.1$,
- absorption coefficient of the absorber in visible region = 0.935,
- thermal conductivity of insulation material 1 (Figure 16) = 0.02 W/m.K,
- thermal conductivity for insulation material 2 = 0.1 W/m.K [63].

Zhai H et al. [65] studied a triangle cavity at a lower temperature (< 200 °C). The optical efficiency simulated by using a light tracking method and the thermal performance was tested experimentally under temperature levels of 90 °C and 150 °C [65]. This study found that the triangle shape cavity receiver optical efficiency was 99%, and thermal losses were 20 W and 41 W (measured at 0.5 m of the receiver length) at the inlet temperature of 90 °C and 150 °C, respectively. Moreover, the solar conversion efficiency could be beyond 67% for the triangle cavity [65].

**D. Trapezoidal cavity receiver**

Singh et al. [66][67] studied the effects of various design parameters of a trapezoidal cavity absorber on the thermal performance. The trapezoidal cavity absorber with a round pipes is shown in Figure 18. The absorber pipes were made of six mild steel round tubes brazed together in a single layer. The absorber pipe was at the upper portion of the cavity. Glass wool insulation was provided at the top and the sides of the pipes. At sidewalls of the cavity, ceramic tiles plates were provided. At the bottom part of the cavity, a glass plane was provided as a window for transmitting the solar radiation. Also, this study proposed another trapezoidal cavity receiver with a rectangular pipe absorber instead of round pipes to compare and evaluate their performance, see Figure 19.

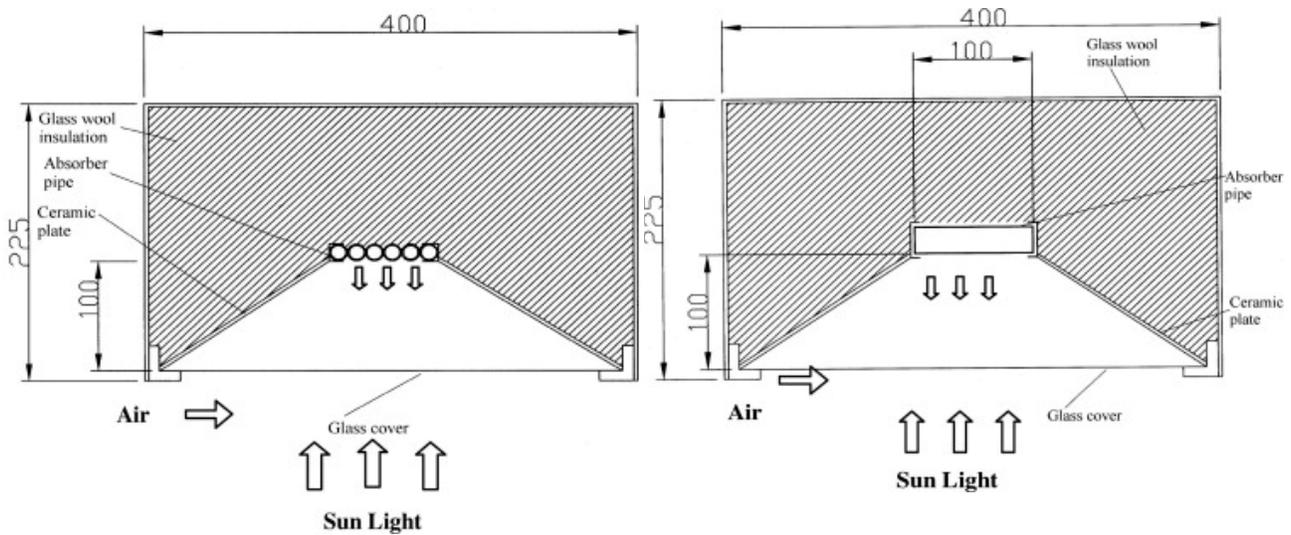

Figure 18: Cross-sectional of the trapezoidal cavity with round pipe absorber [66].

Figure 19: Cross-sectional of the trapezoidal cavity with rectangular pipe absorber [66].

The dimensions of the designs were mentioned in detail in [66]. The thermal performance was measured for eight sets of identical designs with round pipe and rectangular pipe absorbers. The trapezoidal cavity with round pipe absorbers was tested in four different setups. The first two, the round pipe absorbers, were painted with ordinary matt black paint and black nickel coating (selective coating) with emissivity 0.91 and 0.17 at 100 °C, respectively. The other two, the cavity was fabricated with double (10 mm spacing) and single glass cover. Similar to the above mentioned scenarios, the trapezoidal cavity with the rectangular pipe absorbers was tested with four different setups. The experimental results showed that the difference between the heat loss coefficient of rectangular and round pipe absorbers in the trapezoidal cavity were not significantly different – they differed by a factor of 3.3 to 8.2 W/m$^2$, respectively [66]. The selective coating on the absorbers had a remarkable reduction of overall heat loss coefficient by 20% to 30 % compared to ordinary black paint. In addition, using double glass cover reduced the overall heat loss by 10% to 15 % compared to single glass cover [66]. Manikumar et al. [68] analyzed trapezoidal cavity numerically and experimentally. The cavity had a multi-tube absorber with a plate and without plate underneath, see Figure 20 and Figure 21, for various values of gaps between the tubes and depths of the cavity. The values of the overall heat loss coefficient and convective heat transfer coefficient were observed to increase with gaps between the tubes and the tube temperature. The thermal efficiency of the cavity with a plate was higher than without.

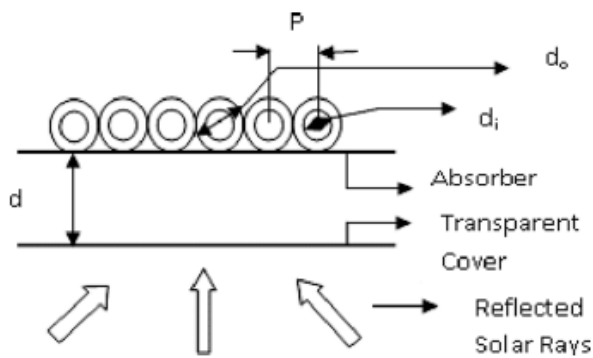
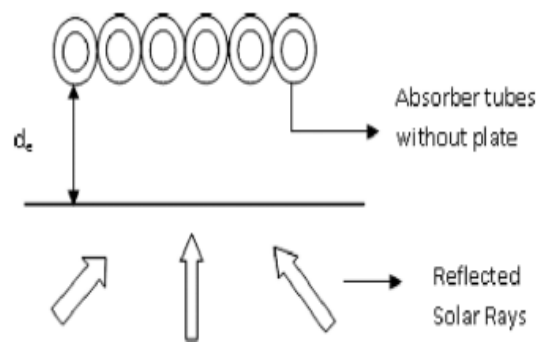

Figure 20: The arrangement of multi-tube absorber with plate[68].

Figure 21: The arrangement of multi-tube absorber with plate [68].

Oliveira et al. [69] analyzed and optimized a trapezoidal cavity receiver via ray-trace and computational fluid dynamic (CFD) simulations. It was found that the cavity with six absorber tubes of 1/2" / 5/8" inner/outer diameters collects had higher optical efficiency. Further, the maximum inclination of 50° of the lateral cavity wall with respect to the bottom base was found to be optically acceptable. CFD simulation was used to optimize the cavity depth and rock wool insulation thickness. The lowest heat transfer coefficient was observed at the cavity depth of 45 mm. The insulation thickness of 35 mm of rock wool showed a good compromise between shading and insulations.

Reynold et al. [70] studied the heat transfer rate and the heat loss of the trapezoidal cavity experimentally with the aim of optimizing the cavity design to achieve maximum thermal efficiency and to develop a numerical model. Furthermore, the cavity was modeled by using a computational fluid dynamic (CFD) software package. The heat transfer rate results and flow patterns showed reasonable agreement between the computational and experimental works, but the heat loss that was measured by CFD was underestimated by about 40% as compared to the experimental results. This discrepancy could not be explained. The heat transfer rate was compared between a uniform and non-uniform heating of the bottom wall with natural convection flows in a trapezoidal cavity using a finite element analysis with bi-quadratic elements (a method for evaluating the nonlinear coupled partial differential equations for flow and temperature fields) [71]. It was found that for all Raleigh numbers, the non-uniform heating of the bottom wall has a significant heat transfer rate as compared to uniform heating [71].

Moreover, the trapezoidal cavity receiver has been studied optically by Liang et al. [72]. The optical efficiency (it is the ratio between the solar radiation reaching the receiver absorber and the solar radiation coming from the concentrator mirror) of the total configuration was around 85% while the optical efficiency of the absorber tube was about 45%.

Natarajan et al. [73] studied the effect of the Grashof number, absorber angle, aspect ratio (ratio of width and depth of cavity), surface emissivity and temperature ratio (ratio of the bottom and top surface temperature of the cavity) of the trapezoidal cavity. In their model, radiation and convection heat transfer was included. Consequently, the Grashof number has been included in the Nusselt number correlations. It was found that the effect of Grashof number on the combined heat loss (natural convection and surface radiation) was negligible. Further, the combined Nusselt number on the absorber angle was also negligible. The Nusselt number was decreased by increasing the aspect ratio and the temperature ratio, while an increase was observed in surface emissivity. Beyond the temperature ratio of 0.6 and aspect ratio 2.5, the combined heat loss variation in the cavity was not notable.

## 4. Conclusion

The alternative to the evacuated receiver unit of the Parabolic Trough Collector (PTC) that can overcome its shortages and compete in terms of efficiency, performance, and cost-effective is the objective of the research of the PTC. Cavity receivers have the potential to overcome the shortages of the evacuated receiver. This work is a comprehensive study of different cavities receiver designs.

It was found that the effect of the cavity aperture size and the mirrored cavity reflectivity on the cavity performance are essential, where the smaller the cavity aperture and the higher cavity mirror reflectivity yield higher efficiency. Further, the reflectivity of the cavity mirror enhances the optical performance and solar flux distribution inside the receiver as well as reducing the temperature gradient and thermal stresses due to the effect of the multiple reflections in the cavity annulus. However, the dimension of the cavity aperture alongside with the focal line of the collector should be carefully selected to ensure all the reflected radiation enters the cavity receiver through the aperture. It is necessary to keep the concentration ratio maximum (it could require to change the parabolic trough rim angle) and to minimize inaccuracies in both tracking and directional errors due to the sun shape. To diminish the effect of changing the focal length on the cavity opening and the error in the tracking angle, the cavity aperture could have a mirrored V shape or adding flat plate reflector at the cavity aperture (optical funnel).

The main deficiencies of most of the existing cavity designs are keeping the cavity annulus under atmospheric pressure and leaving the cavity aperture open. This often leads to significant losses. Generally, the thermal losses of the cavity are affected by the following: surface emissivity, working fluid temperature, mass flow rate, and wind velocity. Also, cavity geometry could affects both optical and thermal efficiency. In which case the cylindrical geometry seems to be more effective, especially in terms of thermal and optical flux distribution.

In non-cylindrical geometry cavity receivers, for the case of the elliptical cavity receiver, which the absorber pipe was positioned on the opposite focus from the cavity opening, was prone to deformation due to non-uniform heat flux absorption across the pipe circumference. The triangular and V-shape cavity receivers had the best optical efficiency among the non-cylindrical cavity receivers followed by the elliptical cavity with a flat plate reflector. However, their heat flux distribution of the heating surface were heterogeneous, which could cause thermal stress at a higher temperature. In addition, their thermal performance was comparable to the evacuated receiver at the medium temperature range and the heat loss was equivalent to the evacuated receiver in a windless condition. Moreover, the optical performance was affected by the aperture width, the depth to width and the offset distance from the focus of the triangular shape.

For the trapezoidal cavity receiver, the optical efficiency of the absorber tube was about 45% and for the total configuration was around 85%. The overall heat loss of the trapezoidal cavity aperture with double-glazed cover reduced by up to 15% compared to the single glass cover.

For cylindrical cavity receivers, where there is no vacuum in the annulus, the convection and conduction heat loss in the annulus can be minimized by selecting the optimum annulus dimension between the absorber pipe and the cavity envelope. Although this largely depends on the diameters of the receiver components. The conventional designs that was developed by coating the inner side of the glass cover with a mirrored or hot mirror coatings had a deficiency of being breakable. The receiver unit with hot mirror over the glass cover with a vacuum in the annulus was found to be more effective than the conventional receiver at a higher temperature. The hot mirror receiver could be better alternative to the conventional receiver with improving hot mirror materials for hot mirror applications especially, the transmissivity in the visible region. In the current receiver unit with hot mirror application, the hybrid system is usually favored.

Those receivers that were utilized air as a working fluid had overcome the lower volumetric heat transfer capacity of the air by increasing the heat transfer area. This was done by eliminating the absorber pipe and using corrugate structure inside the cavity wall to allow the air to flow in a sufficient large cross section, but a significant energy power was required for pumping the working fluid.

In cylindrical geometry designs of the absorber pipe for the cavity receiver such as copper pipes, copper annulus and crescent shaped channel had the aims of raising the HTF and reducing the cost of production and maintenance. It was found that, the heat loss of the cavity receiver with the crescent shaped channel with a borosilicate glass on the cavity aperture was less and slower than that of the evacuated receiver at a high temperature (> 400 °C).

The design of the mirrored cavity receiver with a hot mirror application on the aperture showed that it was able to achieve the highest thermal and optical efficiencies, especially if the hot mirror coating has a higher transmission coefficient for the incident solar radiation. Further, this cavity exceeded the HTF temperature ceiling compared to existing alternatives and hence increase the efficiency of the system.

**Acknowledgments**

The authors would like to thank the following entities for their kind support: MERG (Materials for Energy Research Group) and MITP (Mandelstam Institute for Theoretical Research), at the University of the Witwatersrand.

# 5. References

[1] Chaanaoui M, Vaudreuil S, Bounahmidi T. Benchmark of Concentrating Solar Power Plants: Historical, Current and Future Technical and Economic Development. *Procedia Comput Sci* 2016; 83: 782–789.

[2] Chen F, Li M, Hassanien Emam Hassanien R, et al. Study on the optical properties of triangular cavity absorber for parabolic trough solar concentrator. *Int J Photoenergy*; 2015.

[3] Moya EZ. 7 - Parabolic-trough concentrating solar power (CSP) systems. In: Lovegrove K, Stein W (eds) *Concentrating Solar Power Technology*. Woodhead Publishing, pp. 197–239.

[4] Forristall R. *Heat transfer analysis and modeling of a parabolic trough solar receiver implemented in engineering equation solver*. National Renewable Energy Lab., Golden, CO.(US), http://large.stanford.edu/publications/coal/references/troughnet/solarfield/docs/34169.pdf (2003, accessed 13 August 2017).

[5] Ramchandra G. P, Sudhir V. P, Jyeshtharaj B. J, et al. Alternative designs of evacuated receiver for parabolic trough collector. *Energy*; 155: 66–76.

[6] Mohamad K, Ferrer P. Parabolic trough efficiency gain through use of a cavity absorber with a hot mirror. *Appl Energy* 2019; 238: 1250–1257.

[7] Fuqiang W, Ziming C. Progress in concentrated solar power technology with parabolic trough collector system: A comprehensive review. *Renew Sustain Energy Rev* 2017; 79: 1314–1328.

[8] Wang P, Liu DY, Xu C. Numerical study of heat transfer enhancement in the receiver tube of direct steam generation with parabolic trough by inserting metal foams. *Appl Energy* 2013; 102: 449–460.

[9] Sargent, Lundy LLC Consulting Group. *Assessment of Parabolic Trough and Power Tower Solar Technology Cost and Performance Forecasts*. Chicago, Illinois: DIANE Publishing, https://scholar.google.co.za/scholar?hl=en&as_sdt=0%2C5&q=Assessment+of+Parabolic+Trough+and+Power+Tower+Solar+Technology+Cost+and+Performance+Forecasts&btnG= (2003).

[10] International Renewable Energy Agency. Renewable energy technologies: cost analysis series: concentrating solar power; 2012., https://www.irena.org/DocumentDownloads/Publications/RE_Technologies_Cost_Analysis-CSP.pdf (accessed 13 August 2017).

[11] Antonaia A, Castaldo A, Addonizio ML, et al. Stability of W-Al 2 O 3 cermet based solar coating for receiver tube operating at high temperature. *Sol Energy Mater Sol Cells* 2010; 94: 1604–1611.

[12] Dudley VE, Kolb GJ, Mahoney AR, et al. Test results: SEGS LS-2 solar collector. *Nasa Stirecon Tech Rep N*; 96.

[13] Forristall R. EES heat transfer model for solar receiver performance. *Proc ISEC Sol* 2004; 11–14.

[14] Kennedy CE. *Review of mid-to high-temperature solar selective absorber materials*. National Renewable Energy Lab., Golden, CO.(US), http://large.stanford.edu/publications/power/references/troughnet/solarfield/docs/31267.pdf (2002, accessed 8 August 2017).

[15] Kennedy CE, Price H. Progress in Development of High-Temperature Solar-Selective Coating. *ASME Int Sol Energy Conf Sol Energy* 2005; 749–755.

[16] ASE. Archimede Solar Energy, http://www.archimedesolarenergy.it/ (accessed 8 December 2017).

[17] SCHOTT. SCHOTT PTR 70, https://www.schott.com/d/csp/370a8801-3271-4b2a-a3e6-c0b5c78b01ae/1.0/schott_ptr70_4th_generation_brocure.pdf (accessed 27 February 2019).

[18] Reddy KS, Ravi Kumar K, Ajay CS. Experimental investigation of porous disc enhanced receiver for solar parabolic trough collector. *Renew Energy* 2015; 77: 308–319.

[19] Ghasemi S, Ranjbar A. Numerical thermal study on effect of porous rings on performance of solar parabolic trough collector. *Appl Therm Eng* 2017; 118: 807–816.

[20] Mwesigye A, Bello-Ochende T, Meyer JP. Heat transfer and thermodynamic performance of a parabolic trough receiver with centrally placed perforated plate inserts. *Appl Energy* 2014; 136: 989–1003.

[21] Sahin H, Baysal E. Investigation of heat transfer enhancement in a new type heat exchanger using solar parabolic trough systems. *Int J Hydrog Energy* 2015; 40: 15254–15266.

[22] Eiamsa-ard S, Thianpong C, Eiamsa-ard P. Turbulent heat transfer enhancement by counter/co-swirling flow in a tube fitted with twin twisted tapes. *Exp Therm Fluid Sci* 2010; 34: 53–62.

[23] Diwan K, Soni MS. Heat transfer enhancement in absorber tube of parabolic trough concentrators using wire-coils inserts. *Univers J Mech Eng* 2015; 3: 107–112.

[24] Martín RH, Pérez-García J, García A, et al. Simulation of an enhanced flat-plate solar liquid collector with wire-coil insert devices. *Sol Energy* 2011; 85: 455–469.

[25] Shashank SC, Taji SG. Experimental studies on effect of coil wire insert on heat transfer enhancement and friction factor of double pipe heat exchanger. *Int J Comput Eng Res*; 3.

[26] Nanan K, Thianpong C, Pimsarn M, et al. Flow and thermal mechanisms in a heat exchanger tube inserted with twisted cross-baffle turbulators. *Appl Therm Eng* 2017; 114: 130–147.

[27] Hong Y, Du J, Wang S. Experimental heat transfer and flow characteristics in a spiral grooved tube with overlapped large/small twin twisted tapes. *Int J Heat Mass Transf* 2017; 106: 1178–1190.

[28] Huang Z, LI Z. Numerical investigations on fully-developed mixed turbulent convection in dimpled parabolic trough receiver tubes. *Appl Therm Eng* 2017; 114: 1287–1299.

[29] Cheng Z, He YL. Numerical study of heat transfer enhancement by unilateral longitudinal vortex generators inside parabolic trough solar receivers. *Int J Heat Mass Transf* 2012; 55: 5631–5641.

[30] Fuqiang W, Qingzhi L, Huaizhi H. Parabolic trough receiver with corrugated tube for improving heat transfer and thermal deformation characteristics. *Appl Energy* 2016; 164: 411–424.

[31] Fuqiang W, Zhexiang T, Xiangtao G. Heat transfer performance enhancement and thermal strain restrain of tube receiver for parabolic trough solar collector by using asymmetric outward convex corrugated tube. *Energy* 2016; 114: 275–292.

[32] Bellos E, Tzivanidis C, Antonopoulos KA. Thermal enhancement of solar parabolic trough collectors by using nanofluids and converging-diverging absorber tube. *Renew Energy* 2016; 94: 213–222.

[33] Hassani S, Saidur R, Mekhilef S, et al. A new correlation for predicting the thermal conductivity of nanofluids; using dimensional analysis. *Int J Heat Mass Transf* 2015; 90: 121–130.

[34] Yunus CA, Afshin JG. *Heat and mass transfer: fundamentals and applications*. 5 Ed. 2015.

[35]  Patil RG, Panse SV, Joshi JB. Optimization of non-evacuated receiver of solar collector having non-uniform temperature distribution for minimum heat loss. *Energy Convers Manag* 2014; 85: 70–84.

[36]  Stettenheim J, McBride TO, Brambles OJ. *Cavity Receivers for Parabolic Solar Troughs*. 2013.

[37]  Kaluba VS, Ferrer P. A model for hot mirror coating on solar parabolic trough receivers. *J Renew Sustain Energy* 2016; 8: 053703.

[38]  Kaluba VS, Mohamad K, Ferrer P. Experimental and simulated Performance of Heat Mirror Coatings in a Parabolic Trough Receiver. *Submitt Appl Energy J ArXiv190800866 Physicsapp-Ph*.

[39]  Granqvist CG. Spectrally Selective Coatings for Energy Efficiency and Solar Applications. *Phys Scr* 1985; 32: 401.

[40]  Canan K. Performance analysis of a novel concentrating photovoltaic combined system. *Energy Convers Manag* 2013; 67: 186–196.

[41]  Miller DC, Khonkar HI, Herrero R, et al. An end of service life assessment of PMMA lenses from veteran concentrator photovoltaic systems. *Sol Energy Mater Sol Cells* 2017; 167: 7–21.

[42]  Lampert CM. Coatings for enhanced photothermal energy collection II. Non-selective and energy control films. *Sol Energy Mater* 1979; 2: 1–17.

[43]  Grena R. Efficiency Gain of a Solar Trough Collector Due to an IR-Reflective Film on the Non-Irradiated Part of the Receiver. *Int J Green Energy* 2011; 8: 715–733.

[44]  Granqvist CG. Radiative heating and cooling with spectrally selective surfaces. *Appl Opt* 1981; 20: 2606–2615.

[45]  Cyulinyana MC, Ferrer P. Heat efficiency of a solar trough receiver with a hot mirror compared to a selective coating. *South Afr J Sci* 2011; 107: 01–07.

[46]  Mohamad K, Ferrer P. *[Patent] Thermal radiation loss reduction in a parabolic trough receiver by the application of a cavity mirror and a hot mirror coating*. PCT/IB2019/053531, South Africa, Johannesburg.

[47]  Mohamad K, Ferrer P. Experimental and numerical measurement of the thermal performance for parabolic trough solar concentrators. *Proc 63th Annu Conf South Afr Inst Phys SAIP2018 ArXiv190800515 Physicsapp-Ph*.

[48]  Mohamad K, Ferrer P. Computational comparison of a novel cavity absorber for parabolic trough solar concentrators. *Proc 62th Annu Conf South Afr Inst Phys SAIP2017 ArXiv190713066 Physicsapp-Ph*.

[49]  Bader R, Pedretti A, Barbato M, et al. An air-based corrugated cavity-receiver for solar parabolic trough concentrators. *Appl Energy* 2015; 138: 337–345.

[50]  Bader R, Pedretti A, Steinfeld A. Experimental and numerical heat transfer analysis of an air-based cavity-receiver for solar trough concentrators. *J Sol Energy Eng* 2012; 134: 021002.

[51]  Hänchen M, Brückner S, Steinfeld A. High-temperature thermal storage using a packed bed of rocks – Heat transfer analysis and experimental validation. *Appl Therm Eng* 2011; 31: 1798–1806.

[52]  G. Z, A. P, A. H, et al. Design of packed bed thermal energy storage systems for high-temperature industrial process heat. *Appl Energy*; 137: 812–822.

[53]  Burkholder F, Kutscher C. *Heat Loss Testing of Schott's 2008 PTR70 Parabolic Trough Receiver*. NREL/TP--550-45633, 1369635. Epub ahead of print 1 May 2009. DOI: 10.2172/1369635.

[54]   Barra OA, Franceschi L. The parabolic trough plants using black body receivers: experimental and theoretical analyses. *Sol Energy* 1982; 28: 163–171.

[55]   Boyd DA, Gajewski R, Swift R. A cylindrical blackbody solar energy receiver. *Sol Energy* 1976; 18: 395–401.

[56]   Li X, Chang H, Duan C, et al. Thermal performance analysis of a novel linear cavity receiver for parabolic trough solar collectors. *Appl Energy* 2019; 237: 431–439.

[57]   Cao F, Li Y, Wang L, et al. Thermal performance and stress analyses of the cavity receiver tube in the parabolic trough solar collector. In: *IOP Conference Series: Earth and Environmental Science*. IOP Publishing, 2016, p. 012067.

[58]   Fei C, Lei W, Tianyu Z. Design and Optimization of Elliptical Cavity Tube Receivers in the Parabolic Trough Solar Collector. *Int J Photoenergy*; 2017: 7.

[59]   Chen F, Li M, Xu C, et al. Study on heat loss performance of triangle cavity absorber for parabolic trough concentrators. In: *Materials for Renewable Energy and Environment (ICMREE), 2013 International Conference on*. IEEE, 2014, pp. 683–689.

[60]   Xiong Y, Wu Y, Ma C, et al. Numerical investigation of thermal performance of heat loss of parabolic trough receiver. *Sci China Technol Sci* 2010; 53: 444–452.

[61]   Xiao X, Zhang P, Shao DD, et al. Experimental and numerical heat transfer analysis of a V-cavity absorber for linear parabolic trough solar collector. *Energy Convers Manag* 2014; 86: 49–59.

[62]   Chen F, Li M, Zhang P, et al. Thermal performance of a novel linear cavity absorber for parabolic trough solar concentrator. *Energy Convers Manag* 2015; 90: 292–299.

[63]   Hongbo L, Chunguang Z, Man F. Study on the thermal performance of a novel cavity receiver for parabolic trough solar collectors. *Appl Energy* 2018; 222: 790–798.

[64]   Liang H, Fan M, You S, et al. An analysis of the heat loss and overheating protection of a cavity receiver with a novel movable cover for parabolic trough solar collectors. *Energy* 2018; 158: 719–729.

[65]   Zhai H, Dai Y, Wu J, et al. Study on trough receiver for linear concentrating solar collector. In: *Proceedings of ISES World Congress 2007 (Vol. I–Vol. V)*. Springer, 2008, pp. 711–715.

[66]   Singh PL, Sarviya RM, Bhagoria JL. Heat loss study of trapezoidal cavity absorbers for linear solar concentrating collector. *Energy Convers Manag* 2010; 51: 329–337.

[67]   Singh PL, Sarviya RM, Bhagoria JL. Thermal performance of linear Fresnel reflecting solar concentrator with trapezoidal cavity absorbers. *Appl Energy* 2010; 87: 541–550.

[68]   Manikumar R, Palanichamy R, Arasu AV. Heat transfer analysis of an elevated linear absorber with trapezoidal cavity in the linear Fresnel reflector solar concentrator system. *J Therm Sci* 2015; 24: 90–98.

[69]   Facão J, Oliveira AC. Numerical simulation of a trapezoidal cavity receiver for a linear Fresnel solar collector concentrator. *Renew Energy* 2011; 36: 90–96.

[70]   Reynolds DJ, Jance MJ, Behnia M, et al. An experimental and computational study of the heat loss characteristics of a trapezoidal cavity absorber. *Sol Energy* 2004; 76: 229–234.

[71]   Natarajan E, Basak T, Roy S. Natural convection flows in a trapezoidal enclosure with uniform and non-uniform heating of bottom wall. *Int J Heat Mass Transf* 2008; 51: 747–756.

[72]  Liang H, Fan M, You S, et al. A Monte Carlo method and finite volume method coupled optical simulation method for parabolic trough solar collectors. *Appl Energy* 2017; 201: 60–68.

[73]  Natarajan SK, Reddy KS, Mallick TK. Heat loss characteristics of trapezoidal cavity receiver for solar linear concentrating system. *Appl Energy* 2012; 93: 523–531.